\documentclass[conference]{IEEEtran}
\IEEEoverridecommandlockouts

\usepackage{cite}
\usepackage{todonotes}
\usepackage{amsmath,amssymb,amsfonts}
\usepackage{algorithmic}
\usepackage{graphicx}
\usepackage{textcomp}
\usepackage{xcolor}
\usepackage{url} 
\def\BibTeX{{\rm B\kern-.05em{\sc i\kern-.025em b}\kern-.08em
    T\kern-.1667em\lower.7ex\hbox{E}\kern-.125emX}}

\begin{document}

\title{FL-APU: A Software Architecture to Ease Practical Implementation of Cross-Silo Federated Learning\\
\thanks{This work was co-funded by the German Federal Ministry of Education and Research (BMBF) under Grant 13FH587KX1 (FederatedForecasts) and the European Union in the Interreg Upper Rhine programme (project Aura.ai). © © 2024 IEEE. Personal use of this material is permitted. Permission from IEEE must be obtained for all other uses, in any current or future media, including reprinting/republishing this material for advertising or promotional purposes, creating new collective works, for resale or redistribution to servers or lists, or reuse of any copyrighted component of this work in other works. DOI: 10.1109/FLTA63145.2024.10839980}
}

\author{\IEEEauthorblockN{Fabian Stricker}
\IEEEauthorblockA{\textit{Institute of Data-Centric Software Systems} \\
\textit{Karlsruhe University of Applied Sciences}\\
Karlsruhe, Germany \\
fabian.stricker@h-ka.de}
\and
\IEEEauthorblockN{José Antonio Peregrina Pérez}
\IEEEauthorblockA{\textit{Institute of Data-Centric Software Systems} \\
\textit{Karlsruhe University of Applied Sciences}\\
Karlsruhe, Germany \\
jose\_antonio.peregrina\_perez@h-ka.de}
\and
\IEEEauthorblockN{David Bermbach}
\IEEEauthorblockA{\textit{Scalable Software Systems Research Group} \\
\textit{Technische Universität Berlin}\\
Berlin, Germany \\
db@3s.tu-berlin.de}
\and
\IEEEauthorblockN{Christian Zirpins}
\IEEEauthorblockA{\textit{Institute of Data-Centric Software Systems} \\
\textit{Karlsruhe University of Applied Sciences}\\
Karlsruhe, Germany \\
christian.zirpins@h-ka.de}
}

\maketitle

\begin{abstract}
Federated Learning (FL) is an upcoming technology that is increasingly applied in real-world applications. 
Early applications focused on cross-device scenarios, where many participants with limited resources train machine learning (ML) models together, e.g., in the case of Google's GBoard.
Contrarily, cross-silo scenarios have only few participants but with many resources, e.g., in the healthcare domain. 
Despite such early efforts, FL is still rarely used in practice and best practices are, hence, missing. 
For new applications, in our case inter-organizational cross-silo applications, overcoming this lack of role models is a significant challenge.

In order to ease the use of FL in real-world cross-silo applications, we here propose a scenario-based architecture for the practical use of FL in the context of multiple companies collaborating to improve the quality of their ML models. 
The architecture emphasizes the collaboration between the participants and the FL server and extends basic interactions with domain-specific features. First, it combines governance with authentication, creating an environment where only trusted participants can join. Second, it offers traceability of governance decisions and tracking of training processes, which are also crucial in a production environment.
Beyond presenting the architectural design, we analyze requirements for the real-world use of FL and evaluate the architecture with a scenario-based analysis method.
\end{abstract}


\section{Introduction} \label{sec:intro}
State-of-the-art machine learning (ML) models require a large amount of data to achieve good results. 
Sometimes such data, however, is not available locally and can also not be obtained from third parties, e.g., due to data privacy issues or sensitive business secrets~\cite{paper_pallas_fog4privacy}.

In such scenarios, federated learning (FL)~\cite{mcmahanCommunicationEfficientLearningDeep2017} can help by training local models on sensitive data and only sharing the trained models which are then aggregated into a single global model.
As a bleeding-edge technology, FL still faces challenges, e.g., regarding security~\cite{tolpeginDataPoisoningAttacks2020}, privacy~\cite{melis2019exploiting}, or quality~\cite{li2022federated}.
Despite this, FL is already being used in practice for a few application domains, e.g., in Google's GBoard for improving mobile device word completion~\cite{hardFederatedLearningMobile2019}. 
For inter-organizational cross-silo applications, however, public examples are still missing.

One such application is studied in the FederatedForecasts research project\footnote{\url{https://www.h-ka.de/en/idss/projects/federatedforecasts}} which aims to improve short-term energy forecasting for wind and solar power generation.
For this, it uses FL in a cross-silo scenario with multiple energy providers and a trusted third party as coordinator.

While applying FL methods in FederatedForecasts, we identified several problems:
First, while there is much research on FL fundamentals, e.g.,\cite{li2022federated},\cite{yang2019federated},\cite{pillutla2022robust}, there is a lack of research on implementing real-world FL systems and maintaining them for an extended period of time.
Second, while there are reference architectures for FL, such as~\cite{loFLRAReferenceArchitecture2021}, they focus on generic functionality that can be adapted to many different FL settings.
Thus, they ignore domain-specific details and components that are relevant in the context of inter-organizational cross-silo applications.

In this paper, we add to closing the above mentioned gap through FL-APU, a scenario-based FL architecture focusing on collaboration.
Following the FederatedForecasts scenario, we identify real-world requirements and tailor existing reference architectures for practical industry application.
In this regard, we make the following contributions:

\begin{itemize}
    \item We identify and define requirements for cross-or\-ga\-ni\-za\-tio\-nal FL (Section~\ref{sec:requirements}). 
    \item We define and discuss roles which are necessary for inter-organizational FL architectures (Section~\ref{sec:roles}). 
    \item We describe two reference architectures for FL servers (Sections~\ref{sec:fl_server_architecture}) and clients (Section~\ref{sec:fl_client_architecture}).
    \item We identify and analyze key features such as governance and metadata management (Section~\ref{sec:key_processes}).
    \item We validate our design through a scenario-based evaluation of the architecture (Section~\ref{sec:evaluation}).
\end{itemize}
\section{Related Work} \label{sec:related_work}
This section discusses related work on FL systems architecture and development methods.

Bonawitz et al.~\cite{bonawitzFederatedLearningScale2019} created a system design for a cross-device scenario with billions of devices. 
Handling such numbers of devices leads to several problems, e.g., client connectivity, resource limitations, and scalability in regard to the aggregation and the coordinators that manage different populations of participants.
In addition, Stojkovic et al.\cite{stojkovic2022appliedfederatedlearningarchitectural} also proposed an architecture for a similar problem. While scalability is necessary, they mention additional challenges, such as slow release cycles, low client participation, and bias-free metric calculation.
These issues are less pronounced in cross-silo, cross-organizational scenarios because there are only a few participants with many resources. Furthermore, these participants are less likely to drop out of the process and usually always participate in the training. 
In addition, critical organizational aspects such as metadata management and a platform for collaboration are missing.

Cheng and Long~\cite{chengFederatedLearningOperations2022} propose Federated Learning Operations (FLOps), a methodology to develop FL cross-silo systems. 
In comparison, they provide a methodological perspective and do not focus on architecture design.

Lo et al.~\cite{loFLRAReferenceArchitecture2021} proposed a pattern-oriented reference architecture for general FL. 
Their architecture includes fundamental components (e.g., model trainer, aggregator, evaluator, monitoring, and job creator) that provide a good basis for creating customized FL processes. 
However, we focus on the practical use of FL in cross-silo scenarios, including cooperation and collaboration of multiple companies (e.g., negotiation of the FL process configuration). 
Here, some aspects are missing due to the abstraction level and collection of patterns used, e.g., metadata management, a platform for collaboration, data validation, and reporting.

There are a few companies that already offer FL frameworks, such as IBM\cite{ludwigIBMFederatedLearning2020} and NVIDIA\cite{rothNVIDIAFLAREFederated2022}.
However, frameworks are primarily used as a foundation to ease implementation for building experimental or more extensive applications upon it. Hence, a simple framework without management, monitoring, collaboration, or security can not be used as a production-ready application.

In the case of software, Galtier and Marini\cite{galtier2019substraframeworkprivacypreservingtraceable} proposed Substra, an open-source FL software. Substra uses a decentralized FL architecture focusing on collaboration, traceability, and privacy. It consists of different components, such as an asset network to exchange algorithms and models, but it also uses ledgers to trace the training and results.
Compared to our approach, the collaboration is only focused on data and is missing a platform to negotiate the configuration of the FL process. Consequently, the resulting decisions of the negotiation are not tracked. Furthermore, we focus on centralized FL. Therefore, there is a difference in the communication and components of the architecture.
\section{Federated Learning in Practice} 
\label{sec:requirements}

In the context of FederatedForecasts, we have analyzed the cross-organizational aspects that an FL implementation needs to consider in practice:
First, it needs to preserve privacy to prevent the leakage of valuable information to other companies or third parties.
Second, it has to ensure a fair process where all participants profit and are rewarded for the value of their contributions. Since all participants are competitors, participating in an unfair process could lead to disadvantages in the market.
Third, to break down the result of the FL process and create a transparent process, the decisions and information of the process have to be tracked.
Fourth, the context of the project is energy production, which is considered critical infrastructure.
Hence, the quality and robustness of the predictions are essential factors, as well as the reliability of the process.

Beyond the project context, the companies have additional use cases and requirements for the architecture design.
During the prototyping phase, we identified several problems with the communication between company infrastructures. 
First, each of them is subject to different security policies. 
Second, individual experiences with different communication technologies influence the perception of their suitability. 
Accordingly, we systematically analyzed these areas in individual meetings with the security officers from each involved company, where we discussed the requirements regarding communication of information, security, and privacy.
Besides the security-related concerns, we also discussed how the companies can include their requirement and experience in the FL process to achieve a benefit for all participants. The discussions resulted in the following essential requirements:

\begin{enumerate}
    \item For the communication between FL Server and FL Client, a virtual private network (VPN) solution is ruled out because of potentially malicious network-level access.
    \item Using a remote procedure call (RPC) that does not ensure HTTP2-based communication over the internet is likewise undesired to prevent remote execution of malicious code.
    \item The trained models should be stored and tracked because historic models from earlier training runs could achieve better performance.
    \item A decision-making process is essential so that each company can include its experience with ML models in the training process.
    \item Participating companies require sovereign management of their clients to increase data privacy in the presence of involved market competitors.
    \item An external server is not allowed to send messages that start operations within the company infrastructure.
\end{enumerate}
\section{Architectural Overview and Role Model} \label{sec:roles} 

\begin{figure}
    \centering
    \includegraphics[scale=0.40]{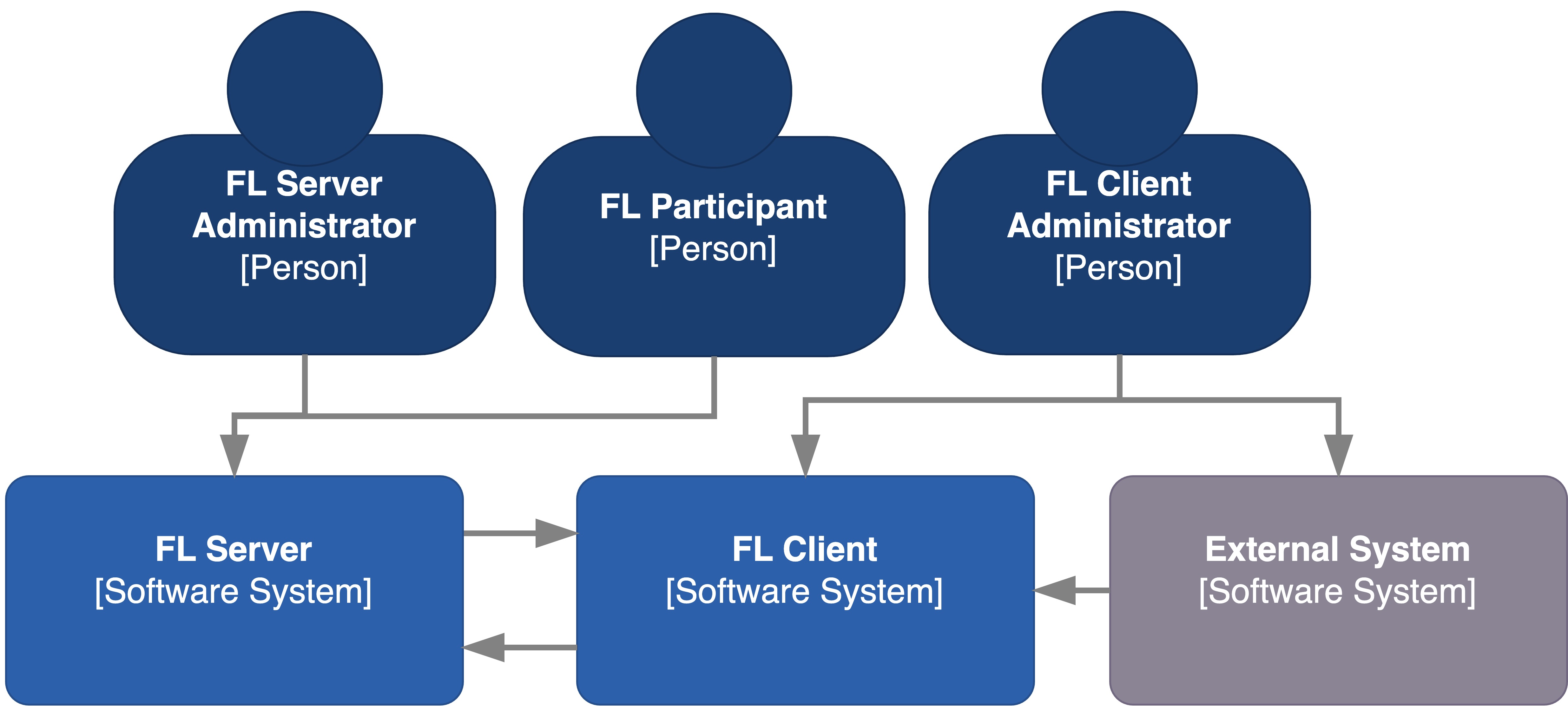}
    \caption{FL-APU: Overview of software systems and their interactions with their surroundings}
    \label{fig:fl_system_view}
\end{figure}

From a practical perspective, the operation of a distributed FL system can be broken down into task areas related to its usage, monitoring, and maintenance.
While monitoring can be done automatically to some extent, other task areas require the active involvement of different types of staff.
In order to structure and distribute their responsibilities, role definitions can be defined.
In particular, roles related to systems management are important in widely distributed settings such as FL systems.
Figure \ref{fig:fl_system_view} shows the system context of our proposed architecture, including the different roles that interact with the system.

Our proposed architecture comprises two software systems, the FL Server and the FL Client.
Both components interact with each other to coordinate and train an ML model.
Besides the systems, different roles are involved in the task areas of operating the system:

\begin{itemize}
    \item \textbf{FL Server Administrator}: Manages the FL Server and monitors the overall FL process, e.g., can start test runs to validate system functionality.
    \item \textbf{FL Participant}: Participates in the decision process for the FL training and process parameters, e.g., deciding on a data format or hyperparameters.
    \item \textbf{FL Client Administrator}: Manages the FL Client and monitors the process, e.g., defines the threshold for deploying or replacing models.
\end{itemize}

Besides the roles interacting with the system, we also consider an external system interacting with the FL Client.
The goal is to access the deployed ML model as a service for inference and to use the predictions in other products.

\section{FL Server Architecture} \label{sec:fl_server_architecture}

\begin{figure*}[htb]
    \centering
    \includegraphics[scale=0.41]{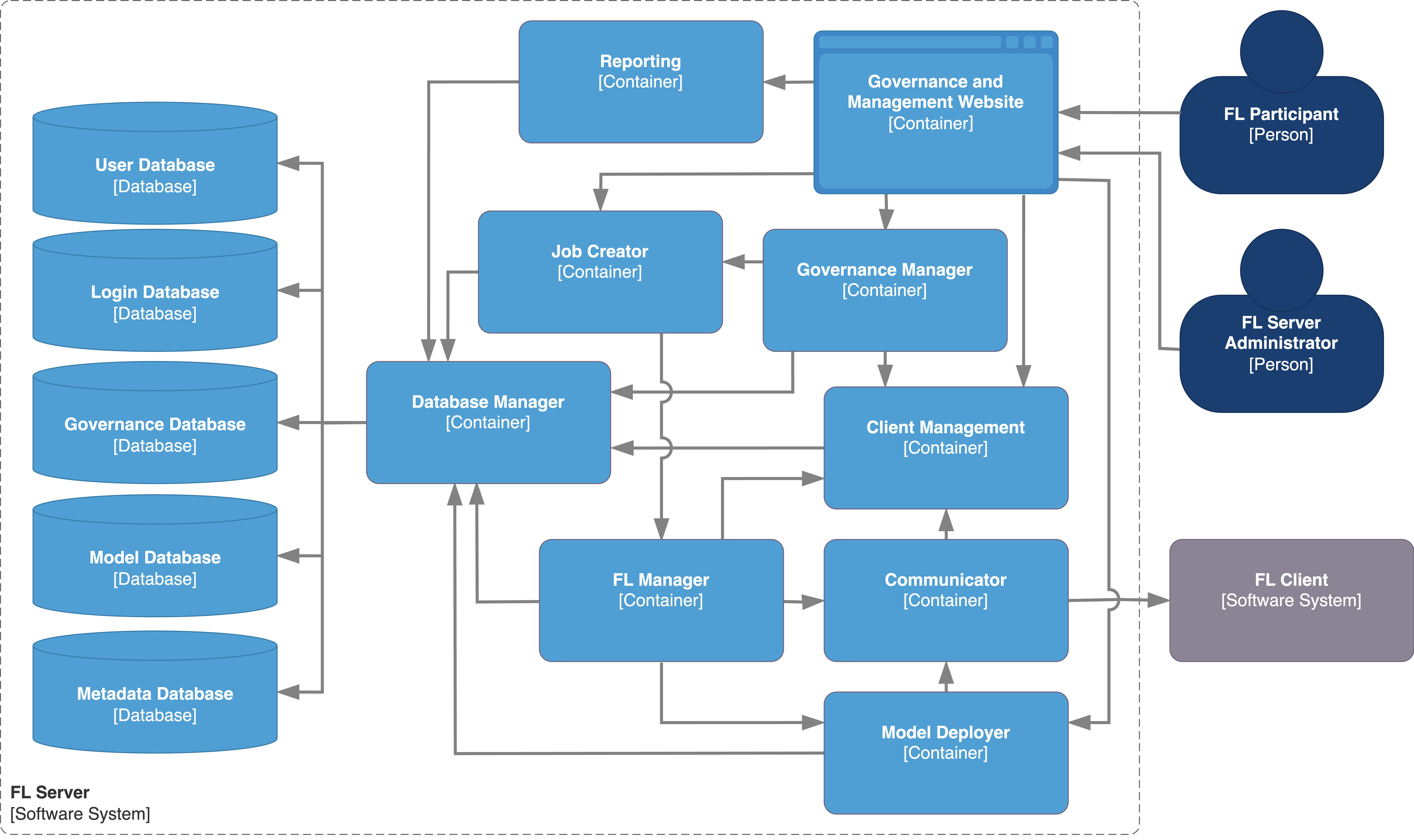}
    \caption{FL-APU: Server architecture as deployable containers; arrows are unidirectional container relationships; containers consist of multiple components}
    \label{fig:fl_server_arch}
\end{figure*}

In this and the following section, we introduce FL-APU, our proposal for an architecture designed for practical use of FL in real-world applications with inter-organizational and cross-silo characteristics.
FL-APU builds on the pattern-oriented reference architecture for federated learning (FLRA)~\cite{loFLRAReferenceArchitecture2021} and refines the architectural design for cross-silo cross-organizational scenarios. 
Furthermore, we extend the base architecture with new components required for the cooperation of organizational participants (i.e., companies) and the integration of the trained model into external applications.
The container view of the FL Server is shown in Figure \ref{fig:fl_server_arch}.
Containers are deployable units that consist of multiple components and focus on fulfilling a specific task in the system. 
Due to the size of the architecture, a visualization on component-level is omitted for the sake of brevity.
We describe the major parts in the following.

\textbf{Governance and Management Website}
The Governance and Management Website enables the FL Server Administrator to access the system and manage the server. 
For example, the administrator can force the deployment of a specific model to a client or perform test runs by manually creating an FL Job to run specific configurations. 
In addition to managing the server, the FL Server Administrator is responsible for registering the FL participants in the Client Management. 
This is needed to ensure that only trusted participants can join the FL process.
Furthermore, the FL Server Administrator and participants can view the current state of the FL process. 
In addition, the FL Participant can take part in the governance negotiation process, which is needed to define the data structure in the data silos, as well as the hyperparameters such as the FL training rounds. 
For example, in our use case, the resolution of the time series data has to be defined. 

\textbf{Governance Manager}
The Governance Manager consists of two components, a \textit{Governance Cockpit} and an \textit{application programming interface} (API).
The Governance Cockpit manages the negotiation process and stores all decisions in a governance contract. 
The contract can be sent to the Job Creator to define an FL Job, including hyperparameters and requirements. 
This is required to allow the participants to include their goals in the process and to collaborate effectively.
We explain the governance aspects of the architecture in Section \ref{sec:key_processes}. 

\textbf{Client Management}
The Client Management consists of three components: \textit{User Management}, \textit{Client Registration}, and \textit{Client Registry}. 
The first one is needed to register the FL participants with a user account and perform authentication of clients. 
The next one is the Client Registration, which accepts registration requests and validates them before they are added to the Client Registry. 
Hence, only legitimate clients can participate in an FL process. 
This container is important for increasing the security of the system, which is essential for the industry scenario.

\textbf{Job Creator}
This container is responsible for creating an FL Job from a governance contract or input from the FL Server Administrator. 
An FL Job contains all parameters required for an FL process, including the training rounds, the train-test-split ratio, evaluation metrics, and more. 
The resulting FL Job is sent to the FL Manager.

\textbf{FL Manager}
The goal of the FL Manager is to handle the whole FL process.
It consists of multiple components, including an \textit{FL Run Manager} that is responsible for managing the other components and starting the process once all required clients are connected to the Client Management. 
In addition, the FL Run Manager can repeat the FL process with different hyperparameters if the participants decide to use hyperparameter optimization.
The managed components are the \textit{Data Validator}, the \textit{Preprocessing Coordinator}, the \textit{Training Coordinator}, the \textit{Model Aggregator}, and the \textit{Evaluation Coordinator}. 
The first one is responsible for validating that the data format decisions are upheld by validating the data structure on the client's side before the training process starts. 
This is important for increasing the robustness of the process by detecting wrong data structures. 
The Preprocessing Coordinator informs the client on how to preprocess the data. 
Next, the Training Coordinator tells the client how to configure the training process.
Similarly, the Evaluation Coordinator informs the client how to perform the evaluation. 
Furthermore, it is also responsible for measuring the client contribution. 
In our scenario, it is important to ensure that each participant contributes equally and that they are compensated based on the value of their contributions. 
The last component is the Model Aggregator, which aggregates client models into the global model.
Generally, the Coordinator components and the Data Validator have a counterpart on the client side.

\textbf{Communicator}
The Communicator manages all the communication between the FL Server and the FL Clients.
It consists of a \textit{Communication Manager} who handles communication with each client and handles encryption and compression of messages. 
Each client has a corresponding Communicator container to receive information from the FL Server.

\textbf{Model Deployer}
Once the training has been completed, the FL Run Manager triggers the Model Deployer to deploy the latest global model on the clients. 
Furthermore, the FL Administrator can deploy a specific model on the clients if an FL Participant requests it.

\textbf{Database Manager}
The Database Manager receives all information regarding users, login, governance, trained models, and metadata. 
This information is stored in the corresponding databases to track the trained model and the overall process.

\textbf{Reporting}
The Reporting component reads the stored information and displays it in a detailed report on the website. 
This is important for preparing a suitable visualization for the FL Participant.

\section{FL Client Architecture} \label{sec:fl_client_architecture}

\begin{figure*}[htb]
    \centering
    \includegraphics[scale=0.40]{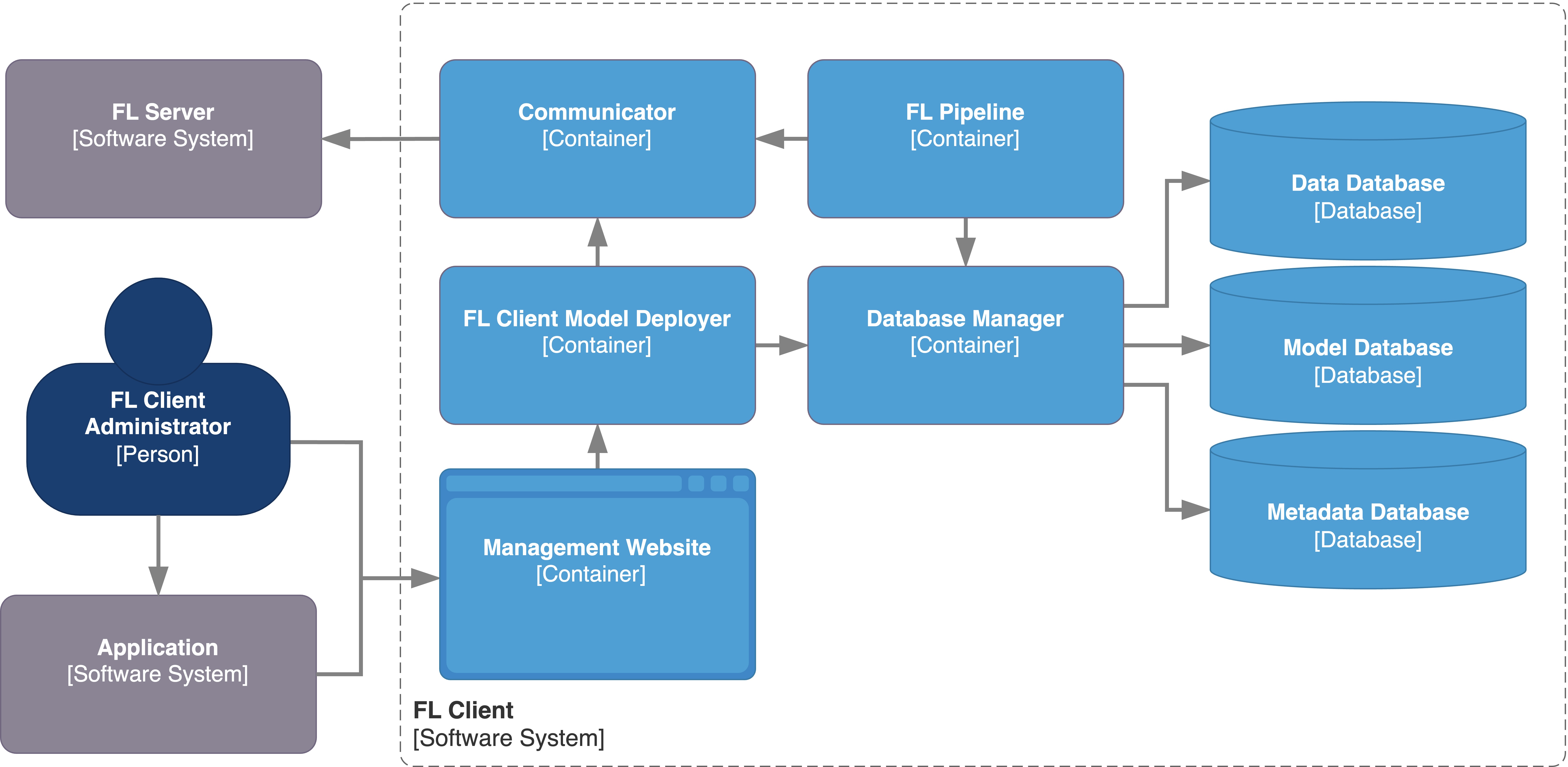}
    \caption{FL-APU: Client architecture as deployable building blocks; Arrows are unidirectional and represent relationships between the containers; containers consist of multiple components}
    \label{fig:fl_client_arch}
\end{figure*}

Since FL always consists of a server and clients, we introduce the architecture for the FL Client as shown in Figure \ref{fig:fl_client_arch}. 
Again, we show the container view of the FL Client and introduce the main containers in the following.

\textbf{Management Website}
The Management Website allows the FL Client Administrator to manage the FL Client.
This includes defining the thresholds for model deployment, defining a model personalization process, monitoring the deployed model, and monitoring the overall FL Client. 
The idea is to enable participating companies to control their FL Clients because these are parts of their individual enterprise IT landscapes. 
Besides the website, this container also includes a \textit{Model Subscription API} component for external systems to perform predictions and use the results in the application. 
Hence, it enables the integration of the FL Client into an external software system.

\textbf{FL Client Model Deployer}
This container consists of five components. 
The first one is the \textit{FL Client Manager}, who handles and tracks the deployment process and prepares the information for the Management Website. 
The next component is \textit{Model Personalization}, which is responsible for personalizing the global model received from the FL Client Manager. 
Another component is the \textit{Decision Maker}, which receives the personalized model to validate whether it fulfills the requirements for deployment. 
If the model fulfills the requirements, it is deployed to the next component, the \textit{Inference Manager}. 
The Inference Manager handles the request from the Model Subscription API and performs the inference on the trained model. 
The last component is \textit{Model Monitoring}, which uses a fixed test dataset to evaluate the performance of the deployed model. 
Once the model reaches a defined threshold or a model is not deployed, the FL Client Manager notifies the FL Client Administrator. 
The administrator can inform the FL Participant to request a different version of the global model or a new training process. 
The validation and monitoring of the deployed model is crucial for a real-world scenario because undetected changes in the model performance can lead to fatal risks.

\textbf{FL Pipeline}
The FL Pipeline consists of one component each for \textit{Data Validation}, \textit{Data Preprocessing}, \textit{Model Trainer}, and \textit{Model Evaluator}. 
These are the coordinator's counterparts on the FL Server. 
The Data Validation executes the configuration to check whether the data structure is valid. 
The next component, Data Preprocessing, takes parameters to execute preprocessing operations. 
The Model Trainer trains the local model using private data and the configuration while the Model Evaluator component uses private test data to validate the model performance of the global model.

\textbf{Communicator}
The Communicator is similar to the one in the FL Server but without managing all client connections.

\textbf{Database Manager}
The Database Manager is also similar to the one in the FL Server. 
In comparison, the client must store the training and evaluation data, client models, and metadata.

\section{Features for Practical FL} \label{sec:key_processes}
In FL-APU, we have different containers to satisfy specific cross-organizational requirements (see Section \ref{sec:requirements}).

In this section, we explain the features of the architecture that aim to tackle these requirements.
While other features are vital, e.g., the model training, we only focus on the features rarely discussed in academia which, however, are essential for the adoption of the architecture in practice.



\textbf{Governance}
In a default setting without governance aspects, the coordinator of the FL system decides how the FL process is conducted.
In this case, the participants only decide if they want to participate. 
For example, the goal of Google's Gboard is to improve the application, and achieve better quality for the clients~\cite{hardFederatedLearningMobile2019}. 
In this case, the one that benefits most is Google because they can offer a better product and may attract more users. 
The participants (i.e., the users) cannot influence the process.
However, in a cross-silo use case, the participating companies expect to profit from the resulting model. 
At the same time, the FL Server is just a necessary component that enables cooperation between the participants and creates the global model but mostly does not benefit from the FL process.
In this regard, the participants must be able to negotiate the FL process configuration to include their input.
For example, the experience with different model architectures or the requirement of explainable models are possible inputs.
Furthermore, the companies will only participate if they receive a benefit.
The goal of Data Governance in FL~\cite{peregrinaDataGovernanceFederated2022} is to provide mechanisms that allow participants to negotiate the training parameters (e.g., the dataset for training and type of model). 
In order to integrate Data Governance into the architecture, two components are required.
The first component is the  Data Governance Cockpit, which contains all the governance mechanisms.
In detail, it provides participants with mechanisms to define, configure, and agree on the training's goals, parameters, and restrictions. 
The managed information to negotiation processes within the scope of the Data Governance lifecycle~\cite{peregrinaDataGovernanceFederated2022}. 
From this agreement, a configuration is created and sent to the FL Manager, which starts the training. 
All operations performed within the Cockpit are recorded as provenance metadata.
The second component is a web interface that provides a GUI for the user to interact with other participants in the decisions.

\textbf{Metadata Management}
The term metadata refers to any data providing information about data. 
Within ML, metadata is used to track ML training experiments and the provenance of the different artifacts involved in the training. 
Hence, it is possible to compare the results achieved by different training runs and the changes that led to either an improvement or deterioration of the model performance. 
As the model performance can depend on many small details, like training data characteristics or hyperparameter tuning, it is necessary to compare the evolution of changes within these details to the model's performance. 
In FL, tracking and comparing these changes is even more important because the performance evaluation is performed by multiple parties, each possessing different data.

Following the work in~\cite{peregrinaMetadataManagementSystem2022a}, this proposal defines a component where metadata are managed.
Provenance metadata allows participants to analyze who performed the operations and the corresponding outcome of the operations.
Experiment tracking metadata provides information on the training results and configuration without ever sharing training data or information on the contents of such data. 
The goal is to prevent the execution of any operation or action that would collide with the concept of FL, where privacy is preserved by design.

\textbf{Data Validation}
A good-performing ML model in FL needs to consider multiple aspects, such as the problem of data heterogeneity, a suitable aggregation method, and model personalization~\cite{li2022federated, tan2022towards}.
In addition, it is vital to have identically structured data on the clients because it is a requirement for horizontal FL~\cite{yang2019federated}, and homogeneous model architecture requires the same input size.

In our architecture, the properties of the data are negotiated during the governance process to solve this issue.
For a more robust FL process, we need to validate that all FL Clients use the correct data structure and that the values are within valid ranges. 
For example, the frequency in a time series dataset should be the same for all FL Clients.
If the data validation fails on a client, the FL Run Manager will identify the client through the Client Management and pause the process. 
The information is stored and reported on the website to inform the FL Participant. 
A similar process is performed on the FL Client to notify the FL Client Administrator.

\textbf{User Authentication}
In FL, preventing unknown clients from joining the training process is essential because such a client could try to steal the global model~\cite{fraboni2021free} or disturb the overall training process with poisoned data~\cite{tolpeginDataPoisoningAttacks2020, zhang2019poisoning}.

In a cross-silo scenario, however, there is only a small number of clients.
Hence, detecting a single malicious client is more realistic than in a cross-device scenario.
Despite this, we still have to ensure that only the participating companies can join the process.
The Client Management container mitigates the problem of unregistered clients by combining the user participating in the negotiation process with the device performing the client-side training. 
Such a process could look like this in practice:

\begin{enumerate}
    \item The participating companies sign a contract with the FL service provider and receive login information for the governance website.
    \item After the governance contract has been completed, each client receives an authentication token for their participating devices.
    \item The participating device is using the token for authentication during the message exchange.
    \item The FL Server is using Client Management to validate the tokens in the request.
\end{enumerate}

With this process, the FL Participant has a separate authentication for the governance website and the participating device.
Furthermore, the token changes after every FL training process. 
Hence, if the token is stolen, the FL Participant can report that their client did not receive a model, and a new process can be started with new tokens.
If the same token is received from two different devices, then the FL Participant could add further information that enables a precise differentiation.
However, the important aspect is that the FL Participant manually takes part in the governance negotiation. 
Consequently, this person can validate that they are still participating in the process and the client is theirs. 
The biggest issue of the process is that once the login information for the governance website is stolen, the company's identity is also stolen.
Thus, a malicious FL Participant can participate in the negotiation process. 
In order to mitigate this problem, the login information should be kept safe, and state-of-the-art security methods (e.g., 2fa and continuous authentication) should be implemented.
In case this happens, the low number of participants would allow the participants to easily restart the entire authentication process, starting from step 2 in the process above.
In addition, we need to consider that multiple FL participants cooperate in the negotiation process. 
Hence, the impact of a single participant wanting to disturb the process is low. 
Furthermore, the configuration has to make sense; otherwise, the participants and the service provider will get suspicious. Finally, authentication is only one part of security, and additional mechanisms are required for more robust security guarantees.

\textbf{Server Authentication and Privacy}
So far, we have only discussed user authentication, but server authentication is also necessary.
Clients connecting to a server must ensure they are dealing with a legitimate server instead of a malicious third party.
A malicious server could try to replace the original one in order to steal client information.
This could be the model updates to perform inference attacks and retrieve private information~\cite{melis2019exploiting}.
Another possibility is that the malicious server could send a manipulated global model.
Such model could achieve good performance but introduce a backdoor to the clients~\cite{bagdasaryan2020backdoor}.
There are state-of-the-art solutions for server authentication that can be implemented in the communicator to validate the genuineness of the server (e.g., certificates).
Furthermore, this is essentially a consensus problem with byzantine failures for which various approaches such as~\cite{kapitza2012cheapbft} exist.


Regarding privacy, the clients should not trust the server even though a contract binds all parties in our scenario. 
Hence, common FL privacy mechanisms such as homomorphic encryption~\cite{park2022privacy} are used in the architecture to increase privacy against leakage of private information from model updates. 
However, ensuring privacy is still an open challenge in FL.
\section{Scenario-based evaluation} \label{sec:evaluation}

To evaluate the architecture design, we follow the scenario-based architecture analysis method (SAAM)~\cite{kazmanScenariobasedAnalysisSoftware1996}.
This method verifies the architecture's design by identifying task scenarios and validating that the architecture can perform these scenarios with the current design. 
In the following, we refer to task scenarios as tasks.
If the architecture does not fulfill a task, the method considers the changes needed to fulfill it and the resulting cost of the changes.
The evaluation method consists of five steps.
In the first step, we describe the architecture, including the containers and their relationships (see Section~\ref{sec:fl_server_architecture} and Section~\ref{sec:fl_client_architecture}).

In the second step, we define tasks that the architecture must fulfill.
In addition to the requirements in Section \ref{sec:requirements} and the system context in Section \ref{sec:roles}, we can deduce the necessary capabilities of an appropriate FL architecture in Table \ref{tab:scenario_tasks}.

\begin{table}[!t]
\caption{Collective tasks for SAAM}
\label{tab:scenario_tasks}
\centering
\begin{tabular}{c||c||c}
\hline
\bfseries ID & \bfseries Actor & \bfseries Task\\
\hline\hline
1 & FL Participant & Participate in the negotiation\\
2 & FL Participant & View FL Run history\\
3 & FL Participant & Request new negotiation process\\
4 & FL Participant & Request deployment of model\\ 
5 & FL Server Admin & Create user accounts \\ 
6 & FL Server Admin & Control the FL process \\
7 & FL Server Admin & Create an FL Job\\
8 & FL Server Admin & Set up a negotiation process\\
9 & FL Client Admin & Set monitoring threshold\\
10 & FL Client Admin & Set deployment threshold\\
11 & FL Client Admin & Monitor the system \\
12 & FL Client Admin & Manage model endpoint \\
13 & FL Server & Prepare a report \\
14 & FL Server & Create a FL Job from Information \\
15 & FL Server & Turn governance result to FL Job \\
16 & FL Server & Store/Retrieve information \\
17 & FL Server & Run FL process \\
18 & FL Server & Deploy a specific model \\
19 & FL Server & Send messages to client \\
20 & FL Server & Encrypt/Compress messages \\
21 & FL Server & Authenticate client \\
22 & FL Server & Generate device token \\
23 & FL Server & Register client \\
24 & FL Server & Monitor FL process \\
25 & FL Server & Check registered clients \\
26 & FL Client & Send messages to server \\
27 & FL Client & Run FL Pipeline \\
28 & FL Client & Store/Retrieve information \\
29 & FL Client & Monitor local FL process \\
30 & FL Client & Configure monitoring \\
31 & FL Client & Configure personalization \\
32 & FL Client & Configure model deployment \\
33 & FL Client & Monitor deployed model \\
34 & FL Client & Encrypt/Compress messages \\
35 & FL Client & Perform model inference \\
36 & FL Client & Perform model personalization \\
37 & FL Client & Decide on model deployment \\
38 & FL Client & Prepare report \\
39 & FL Client & Trigger administrator notification \\
40 & External Application & Send inference request \\ 
\hline
\end{tabular}
\end{table}
Tasks 17 and 27 summarize the typical FL process steps, such as data validation, preprocessing, model training, aggregation of the client models, and evaluation. 


In the third step, we categorize the tasks into direct and indirect ones. 
Direct tasks are supported by the architecture directly, and indirect tasks require changes to the architecture before they can be supported.

In the case of the direct tasks, we analyze Table \ref{tab:scenario_interactions}.
It shows the containers and the tasks they fulfill.
Matching the fulfilled tasks with Table \ref{tab:scenario_tasks}, we can conclude that tasks 1 to 40 are direct tasks that the architecture can execute directly. 


Regarding indirect tasks, the changes in the future are mostly related to different aggregation algorithms, client contribution methods, evaluation methods, or metrics. 
However, the required containers have already been designed to adjust to different methods because these parts can be changed depending on the result of the negotiation and do not require a change in the architecture. 

The requirements of the companies in Section \ref{sec:requirements} guided the overall design of the architecture.
While the first two requirements focus on the actual implementation of the communication, the following three requirements influence the overall design of the process, such as the introduction of a governance component, the tracking of the training runs as well as the management interfaces on the FL Server and the FL Client.
The last requirement influences the communication between the client and the server. 
While coordination and sending messages are vital for the server, the requirement prohibits direct execution of operations. 
Instead of direct communication, where the clients react to messages from the server, the clients have to be proactive in fetching configuration, model weights, and status updates from the server.
In FL-APU, this affects the design of the Communicator. For example, a simple approach could be the implementation of a REST API to store information as resources. The clients periodically retrieve the resources and post client information as a new resource.

In the fourth step, we analyze the interactions with the containers. 
This step is usually done only for the indirect tasks and their changes. 
Since we do not have indirect tasks, we analyze the direct tasks to validate the separation of concerns and which containers fulfill which task.
The containers and their tasks are shown in Table \ref{tab:scenario_interactions}.

\begin{table}[!t]
\caption{List of containers and their associated tasks}
\label{tab:scenario_interactions}
\centering
\begin{tabular}{c||c}
\hline
\bfseries Container & \bfseries Tasks\\
\hline\hline
Reporting & 2,13 \\
Governance and Management Website & 1-8\\
Job Creator & 7,14,15\\
Governance Manager & 3,15\\
Client Management & 5,21,22,25\\
Database Manager & 16\\
FL Manager & 17,24,25 \\
Communicator & 19,20,21,23\\
Model Deployer & 18 \\
\hline
FL Pipeline & 27\\
Management Website & 9,10,11,12,39,40\\
Database Manager & 28\\
FL Client Model Deployer & 9-12,29-33,35-39\\
Communicator & 26,34\\
\hline
\end{tabular}
\end{table}

As a result, several containers execute many tasks, such as the Management Website, FL Client Model Deployer, and the Governance Website. 
However, the containers have a higher abstraction level than the components. 
These containers include components that further distribute these tasks. 
In conclusion, the tasks are sufficiently split on the container level to fulfill the separation of concerns.

In the fifth step, we rank the importance of all tasks.
While most tasks provide essential functionality to the overall FL Process, scenarios 7 and 14 are the least important because they focus on testing a specific FL Run.
All the other tasks are essential and must be implemented to use the architecture.
For example, tracking the process and generating reports without storing information such as metadata is complex.
Another example is the encryption of messages; without it, the communication is insecure and would be rejected by the companies.

As a conclusion of SAAM, the architecture design can fulfill the required tasks; therefore, the design suffices for the overall use case at hand.
We must, however, consider that SAAM only evaluates the architecture design in the context of the predicted usage and cannot anticipate usage scenarios that deviate from the intended path.
\section{Conclusion} \label{sec:conclusion}
Due to the lack of research on implementing and maintaining real-world FL systems for an extended period, realizing an architecture is challenging.
In addition, existing reference architectures primarily focus on generic functionality.
Thus, they ignore domain-specific details and relevant components in the context of inter-organizational cross-silo applications.

We proposed an architecture design for the real-world use of FL in a cross-silo scenario with multiple companies.
The companies aim to train a model collaboratively and use the resulting model in critical infrastructure.
To enable collaboration, we introduced containers that allow the participants to include their input in the configuration of the FL process and decide on a suitable process.
Furthermore, we introduced a container for managing the users and their devices for authentication in order to include only trusted clients in the process.
In addition, the architecture is designed to store and track all decisions and information about the process.
Besides the architecture, we analyzed the requirements of the companies and included them in our scenario-based evaluation. 
In the evaluation, we analyzed how the architecture accommodates the application's context and future changes.
Following the evaluation results, we showed that the architecture design is suitable for our real-world application.
We do not claim that the architecture is completed. 
It will be further improved to adjust to an extended range of scenarios and specific cases, including multiple real-world industrial field studies. 

\bibliographystyle{./IEEEtran}
\bibliography{bibliography}

\end{document}